\documentclass[aps,prd,showpacs,eqsecnum,twocolumn,superscriptaddress]
{revtex4-1}
\usepackage{amsmath,amssymb,graphicx,color}
\usepackage{mathrsfs}

\begin{document}
%\thispagestyle{empty}

%\title{Phenomenological formulation of general relativistic viscous
%  hydrodynamics}

\title{Gravitational waves from remnant massive neutron stars
  of binary neutron star merger: Viscous hydrodynamics effects}

\author{Masaru Shibata}
\affiliation{Center of Gravitational Physics, 
Yukawa Institute for Theoretical Physics, 
Kyoto University, Kyoto, 606-8502, Japan} 

\author{Kenta Kiuchi}
\affiliation{Center of Gravitational Physics, 
Yukawa Institute for Theoretical Physics, 
Kyoto University, Kyoto, 606-8502, Japan} 

\date{\today}
\newcommand{\beq}{\begin{equation}}
\newcommand{\eeq}{\end{equation}}
\newcommand{\beqn}{\begin{eqnarray}}
\newcommand{\eeqn}{\end{eqnarray}}
\newcommand{\pa}{\partial}
\newcommand{\vp}{\varphi}
\newcommand{\varep}{\varepsilon}
\newcommand{\ep}{\epsilon}
\newcommand{\comp}{(M/R)_\infty}
%%%%%%%%%%%%%%%%
\begin{abstract}
%%%%%%%%%%%%%%%%
Employing a simplified version of the Israel-Stewart formalism of
general-relativistic shear-viscous hydrodynamics, we explore the
evolution of a remnant massive neutron star of binary neutron star
merger and pay special attention to the resulting gravitational
waveforms.  We find that for the plausible values of the so-called
viscous alpha parameter of the order $10^{-2}$, the degree of the
differential rotation in the remnant massive neutron star is
significantly reduced in the viscous timescale, $\alt 5$\,ms.
Associated with this, the degree of non-axisymmetric deformation is
also reduced quickly, and as a consequence, the amplitude of
quasi-periodic gravitational waves emitted also decays in the viscous
timescale.  Our results indicate that for modeling the evolution of
the merger remnants of binary neutron stars, we would have to take
into account magnetohydrodynamics effects, which in nature could provide 
the viscous effects.
%%%%%%%%%%%%%%
\end{abstract}
%%%%%%%%%%%%%%


\maketitle

\section{Introduction}

The merger of binary neutron stars is one of the promising sources of
gravitational waves for ground-based gravitational-wave detectors such
as advanced LIGO, advanced VIRGO, and KAGRA~\cite{detectors}. This
fact has motivated the community of numerical relativity to construct
a reliable model for the inspiral, merger, and post-merger stages of
binary neutron stars by numerical-relativity simulations.

The recent discoveries of two-solar mass neutron stars~\cite{twosolar}
strongly constrain the possible evolution processes in the merger and
post-merger stages, because the presence of these heavy neutron stars
implies that the equation of state of neutron stars has to be stiff
enough to support the self-gravity of the neutron stars with mass
$\agt 2M_\odot$. Numerical-relativity simulations with a variety of
hypothetical equations of state that can reproduce the two-solar-mass
neutron stars have shown that massive neutron stars surrounded by a
massive torus are likely to be the canonical remnants formed after the
merger of binary neutron stars of typical total mass
2.6--$2.7M_\odot$~(see, e.g.,
Refs.~\cite{STU2005,Hotokezaka2011,bauswein,Hotokezaka2013,takami,tim}, 
and Refs.~\cite{ms16,luca} for a review).

Because the remnant massive neutron stars are rapidly rotating and
have a non-axisymmetric structure, they could be a strong emitter of
high-frequency gravitational waves of frequency 2.0--3.5\,kHz (e.g.,
Refs.~\cite{bauswein,Hotokezaka2013,takami,tim,ms16,luca}).  A special
attention has been attracted by this aspect in the last decade, and
many {\em purely} hydrodynamics or radiation hydrodynamics simulations
have been performed in numerical relativity for predicting the
gravitational waveforms. However, the remnant massive neutron stars
and a torus surrounding them should be strongly magnetized because
during the merger process, the Kelvin-Helmholtz instability inevitably
occurs and contributes to significantly enhancing the magnetic-field
strength in a timescale much shorter than the dynamical one $\alt
1$\,ms~\cite{PR,kiuchi}. In addition, the remnant massive neutron
stars are in general differentially rotating, and hence, the magnetic
field may be further amplified through magnetorotational
instability~\cite{BH98}.  As a result of these instabilities,
magnetohydrodynamics (MHD) turbulence shall develop as shown by a
number of high-resolution MHD simulations for accretion disks (see,
e.g., Refs.~\cite{alphamodel,suzuki,local}), and it is likely to
determine the evolution of the system in the following manner: (i)
angular momentum would be quickly redistributed; (ii) the degree of
the differential rotation would be quickly reduced; (iii) as a
consequence of the evolution of the rotational velocity profile,
non-axisymmetric deformation may be modified.  By these effects, the
remnant massive neutron stars may result in a weak emitter of 
high-frequency gravitational waves. 

For rigorously exploring these processes resulting from a turbulence
state, extremely-high-resolution MHD simulation is necessary (see,
e.g., Ref.~\cite{kiuchi} for an effort on this).  However, such
simulations will not be practically feasible at least in the next
several years because of the limitation of the computational
resources.  One phenomenological approach for exploring the evolution
of the system that is subject to angular-momentum transport is to
employ viscous hydrodynamics in general
relativity~\cite{DLSS04,radice,SKS16}. The viscosity is likely to be
induced effectively by the local MHD
turbulence~\cite{alphamodel,suzuki,local}, and thus, relying on
viscous hydrodynamics implies that local MHD and turbulence processes
are coarse grained and effectively taken into account. The merit in
this approach is that we do not have to perform an
extremely-high-resolution simulation and we can save the
computational costs significantly.

In this paper, we perform a viscous hydrodynamics simulation for a
remnant massive neutron star of binary neutron star merger.  Following
our previous work~\cite{SKS16}, we employ the Israel-Stewart-type
formalism~\cite{Israel1976}, in which the resulting viscous 
hydrodynamics equations are not parabolic-type but telegraph-type, and 
hence, the causality is preserved~\cite{hiscock}, by contrast to the
cases in which Navier-Stokes-type equations~\cite{LL59} are
employed. In Ref.~\cite{SKS16}, we show that with our choice of the
viscous hydrodynamics formalism, it is feasible to perform long-term
stable simulations for strongly self-gravitating systems. 

This paper is organized as follows: In Sec.~II, we briefly review our
formulation for shear-viscous hydrodynamics.  After we briefly 
describe our setting for numerical simulations in Sec.~III, we present
the results of viscous hydrodynamics evolution for the remnant of
binary neutron star merger in Sec.~IV.  Section~V is devoted to a
summary.  Throughout this paper, $c$ denotes the speed of light. 

%we employ the units of $c=1=G$ where
%$c$ and $G$ denote the speed of light and gravitational constant,
%respectively. 

\section{Formulation for viscous hydrodynamics}

\subsection{Basic equations}

We write the stress-energy tensor of a viscous fluid as~\cite{SKS16}
\beqn T_{ab}=\rho h u_a u_b +P
g_{ab}- \rho h c^{-2} \nu \tau_{ab}^0\, ,
\eeqn
where $\rho$ is the rest-mass density, $h$ is the specific enthalpy,
$u^a$ is the four velocity, $P$ is the pressure, $g_{ab}$ is the
spacetime metric, $\nu$ is the viscous coefficient for the shear
stress, and $\tau^0_{ab}$ is the viscous tensor. Here, $\rho$ obeys
the continuity equation $\nabla_a (\rho u^a)=0$.  In terms of the
specific energy $\varep$ and pressure $P$, $h$ is written as
$h=c^2+\varep+P/\rho$.  $\tau_{ab}^0$ is the symmetric tensor and
satisfies $\tau_{ab}^0 u^a=0$. We suppose that $\nu$ is a function of
$\rho$, $\varep$, $P$, and angular velocity $\Omega$, and will give
the relation below.

Taking into account the prescription of Ref.~\cite{Israel1976}, 
we define that $\tau_{ab}^0$ obeys the following evolution equation:
\beqn
{\cal{L}}_u \tau_{ab}^0=-\zeta (\tau_{ab}^0-\sigma_{ab}),
\eeqn
where ${\cal{L}}_u$ denotes the Lie derivative with respect to $u^a$,
and we write $\sigma_{ab}$ as
\beqn
\sigma_{ab}:=h_a^{~c} h_b^{~d} (\nabla_c u_d + \nabla_d u_c)={\cal{L}}_u h_{ab},
\label{eq:tau}
\eeqn
with $h_{ab}=g_{ab}+u_a u_b$ and $\nabla_a$ the covariant derivative
associated with $g_{ab}$. $\zeta$ is a constant of (time)$^{-1}$
dimension, which has to be chosen in an appropriate manner so that 
$\tau_{ab}^0$ approaches $\sigma_{ab}$ in a short timescale: We
typically choose it so that $\zeta^{-1}$ is shorter than the dynamical
timescale of given systems (but it should be much longer than the
time-step interval of numerical simulations, $\varDelta t$). 

Equation~(\ref{eq:tau}) can be rewritten as
\beqn
{\cal{L}}_u \tau_{ab}=-\zeta \tau_{ab}^0, \label{eq:tauab}
\eeqn
where $\tau_{ab}:=\tau_{ab}^0 - \zeta h_{ab}$. We employ this equation
for $\tau_{ab}$ as one of the basic equations of viscous
hydrodynamics, and hence, the stress-energy tensor is rewritten as
follows:
\beqn
T_{ab}&=&\rho h (1-c^{-2}\nu\zeta) u_a u_b
+ (P -\rho h c^{-2} \nu \zeta)g_{ab}
\nonumber \\
&&- \rho h c^{-2} \nu \tau_{ab}\, . 
\eeqn
Then, the viscous hydrodynamics equations are derived from
$\nabla_a T^a_{~b}=0$ (see Ref.~\cite{SKS16} for details). 

\subsection{Setting viscous parameter}

In the so-called $\alpha$-viscous model, $\nu$ is written as
(see, e.g., Ref.~\cite{ST})
\beqn
\nu=\alpha_v c_s^2 \Omega^{-1}, \label{nunu0}
\eeqn
where $c_s$ is the sound velocity and $\Omega$ denotes the typical
value of the angular velocity.  $\alpha_v$ is the so-called
$\alpha$-viscous parameter, which is a dimensionless constant supposed
to be of the order $10^{-2}$ or more as suggested by latest
high-resolution MHD simulations for accretion
disks~\cite{alphamodel,suzuki,local}.

In this paper, we consider viscous hydrodynamics evolution of a weakly
differentially rotating neutron star, and hence, we set
\beqn
\nu=\alpha_v c_s^2 \Omega_{\rm f}^{-1},\label{nunu}
\eeqn
where we set $\Omega_{\rm f}=2\pi/(0.5\,{\rm ms})$ which is close to
the maximum value of the angular velocity for the initial state of
remnant massive neutron stars (see, e.g., Fig.~\ref{fig2}). We note
that $c_s \Omega_{\rm f}^{-1} \alt 15$\,km with this setting in our
present model.  Since $c_s \Omega_{\rm f}^{-1}$ should not exceed the
size of the neutron star, this is a reasonable setting.

%For the model that we employ in this paper, the relation, $\Omega \alt
%\Omega_e$, is satisfied, and thus, Eq.~(\ref{nunu}) agrees
%approximately with Eq.~(\ref{nunu0}).

The viscous angular momentum transport timescale is approximately
defined by $R^2/\nu$ and estimated to be
\beqn
t_{\rm vis}&\approx & 4.4\,{\rm ms} \left({\alpha_v \over 0.01}\right)^{-1}
\left({c_s \over 0.5c}\right)^{-2}
\left({R \over 10\,{\rm km}}\right)^{2} \nonumber \\
&& \hskip 3cm \times \left({\Omega \over 10^4\,{\rm rad/s}}\right), 
\label{eq3.5}
\eeqn
where we assumed Eq.~(\ref{nunu0}) for $\nu$. We note that for the
central region of the merger remnant, the sound velocity is quite high
typically as $c_s \sim 0.5c$ with maximum $\sim 0.6c$ (and hence, the
velocity in most of the region of the merger remnant is subsonic).  In
the vicinity of the rotation axis (for a small value of $R$), the
timescale may be even shorter.  Thus, within $\sim 5
(\alpha_v/0.01)^{-1}$\,ms, the angular momentum is expected to be
redistributed in the merger remnant.  In the following, we will show
that this is indeed the case. 

\section{Setting for numerical simulations}

\begin{table}[t]
\caption{Key quantities for the initial condition employed in the
  present viscous hydrodynamics simulations: Baryon rest mass, $M_*$,
  gravitational mass, $M$, the maximum rest-mass density, $\rho_{\rm
    max}$, angular velocity at $R=0$, $\Omega_a$, maximum angular
  velocity, $\Omega_{\rm max}$, and dimensionless angular momentum,
  $J/M^2$. The mass, density, and angular velocity are shown in units
  of $M_\odot$, ${\rm g/cm}^3$, and rad/s, respectively.
\label{table1}
}
\begin{center}
\begin{tabular}{cccccc} \hline
  ~$M_*$~&~$M$~
  %&~$R_e$\,(km)~&~$R_c$\,(km)~
& $\rho_{\rm max}$ 
& ~$\Omega_a$~ & ~$\Omega_{\rm max}$~ 
& ~~$J/M^2$~~\\ \hline
~2.94~ & ~2.62~ & $1.0 \times 10^{15}$ & ~$3.2 \times 10^3$~
& ~$9.4 \times 10^3$ &~ 0.76 \\ 
 \hline \hline
\end{tabular}
\end{center}
\end{table}

\subsection{Brief summary of simulation setting}

For solving Einstein's evolution equation, we employ the original
version of Baumgarte-Shapiro-Shibata-Nakamura formulation with a
puncture gauge~\cite{BSSN}.  The gravitational field equations are
solved in the usual fourth-order finite differencing scheme
(e.g., Ref.~\cite{ms16} for a review).

The initial condition of the present simulations is imported from a
simulation result for binary neutron star mergers.  That is, we first
performed a purely hydrodynamics simulation for the merger up to $\sim
5$\,ms after the onset of the merger. The merger simulation was
performed with a grid resolution, which is the same as in the
high-resolution viscous hydrodynamics simulation (see below).  Then,
we extract the weighted rest-mass density, $\rho_*:=\rho \alpha u^t
\psi^6$, weighted spatial velocity field, $\hat u_i:=h u_i$, and
specific energy, $e:=h \alpha u^t-P/(\rho \alpha u^t)$, where $\alpha$
is the lapse function and $\psi$ is the conformal factor for the 
three-dimensional metric, $\psi=[{\rm det}(\gamma_{ij})]^{1/12}$ with
$\gamma_{ij}$ the three-dimensional metric.  We prepare these
quantities in a new computational domain for viscous hydrodynamics and
then solve initial value equations (constraint equations) assuming the
conformal flatness of the system. If necessary, an interpolation for
data is performed (this is the case for lower-resolution runs).  The
assumption of the conformal flatness is acceptable because the
magnitude for all the components of $\psi^{-4}\gamma_{ij}-\delta_{ij}$
is much smaller than unity (at most 0.02) for the merger remnant
employed in this paper. For the initial condition, we set
$\tau^0_{ab}=0$ for simplicity.

%We note that 
%the total rest mass, $M_*$ is derived by
%%%
%\beqn
%M_*&=&\int \rho_* \, d^3x,\\
%J&=& \int \rho_* \hat u_\varphi \, d^3x. 
%\eeqn

In this study, we employ the so-called H4 equation of state~\cite{H4},
approximating it by a piecewise polytropic equation of state with four
pieces~\cite{read}.  During the numerical evolution, we employ a
modified version of the piecewise polytropic equation of state in the
form
\beqn
P=P_{\rm pwp}(\rho)+(\Gamma-1) \rho [\varep-\varep_{\rm pwp}(\rho)],
\eeqn
where $\varep_{\rm pwp}(\rho)$ denotes the specific internal energy
associated with $P_{\rm pwp}$ satisfying $d\varep_{\rm pwp}=-P_{\rm
  pwp}d \rho^{-1}$ and the adiabatic constant $\Gamma$ is set to be
$1.8$. 

Numerical simulations are performed in Cartesian coordinates 
$(x, y, z)$ with a nonuniform grid. 
Specifically, we employ the following grid structure
(the same profile is chosen for $y$ and $z$)
\beqn
\varDelta x=\left\{
\begin{array}{ll}
\varDelta x_0 &~~~ x \leq x_{\rm in}, \\
\varDelta x_i=f\varDelta x_{i-1} &~~~ x > x_{\rm in},
\end{array}
\right.
\eeqn
where $\varDelta x_0$ is the grid spacing in an inner region with
$x_{\rm in} \approx 22.5$\,km.  $\varDelta x_i:= x_{i+1}- x_{i}$ with
$x_i$ the location of $i$-th grid. At $i={\rm in}$, $\varDelta
x_i=\varDelta x_0$.  $f$ determines the nonuniform degree of the grid
spacing and we set it to be $1.03$.  We change $\varDelta x_0$ as
$295$\,m (low resolution), $220$\,m (middle resolution), and $148$\,m
(high resolution) to confirm that the dependence of the numerical
results on the grid resolution is weak (see Appendix A).  The outer
boundary is located at $\approx \pm 2100$\,km along each axis for all
the grid resolutions.  Unless otherwise stated, we will show the
results in the high-resolution runs in the following.  We note that
the wavelength of gravitational waves emitted by remnants of binary
neutron star mergers is typically 120\,km in our present model. Thus,
with this setting of the computational domain, the outer boundary is
located in a wave zone. To suppress the growth of unstable modes
associated with the high-frequency numerical noises, we incorporate a
six-order Kreiss-Oliger-type dissipation term as in our previous
study~\cite{SKS16}.

The viscous coefficient is written in the form of Eq.~(\ref{nunu}).
Latest high-resolution MHD simulations for accretion
disks~\cite{alphamodel,suzuki,local} have suggested that the
$\alpha$-viscous parameter would be of the order $10^{-2}$ or more.
Taking into account this numerical-experimental fact, we choose
$\alpha_v=0.01$ and 0.02. $\zeta$ is set to be $\approx 2\Omega_{\rm
  max} \approx 6\Omega_a$ in this work.

\section{Numerical results}

\begin{figure*}[t]
\begin{center}
\includegraphics[width=56mm]{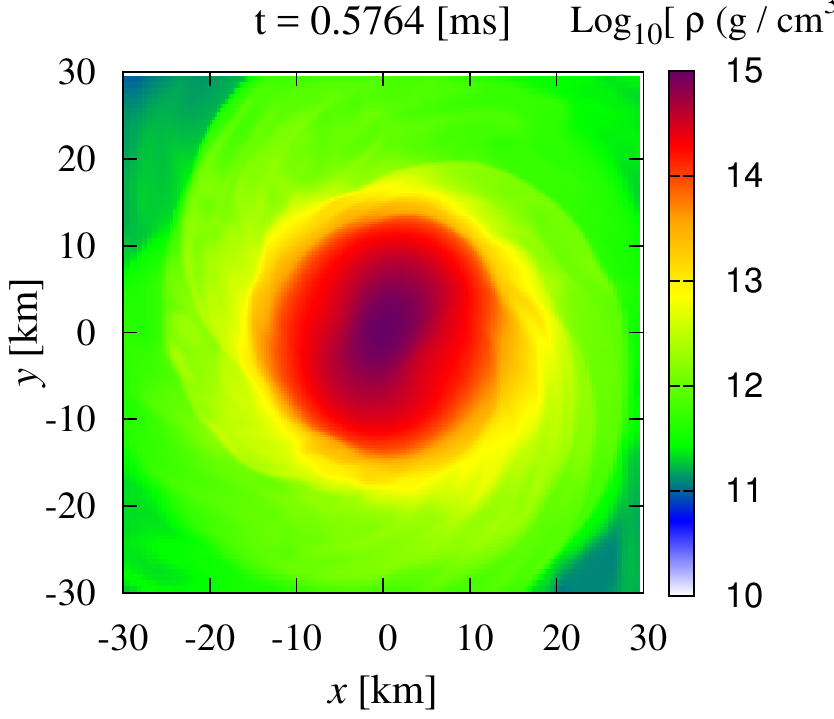}
\includegraphics[width=56mm]{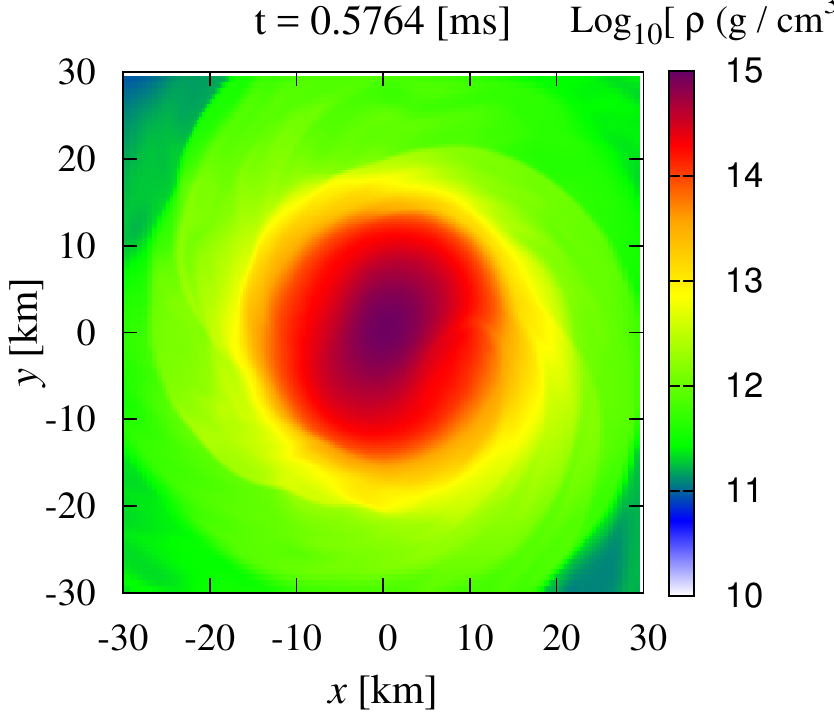}
\includegraphics[width=56mm]{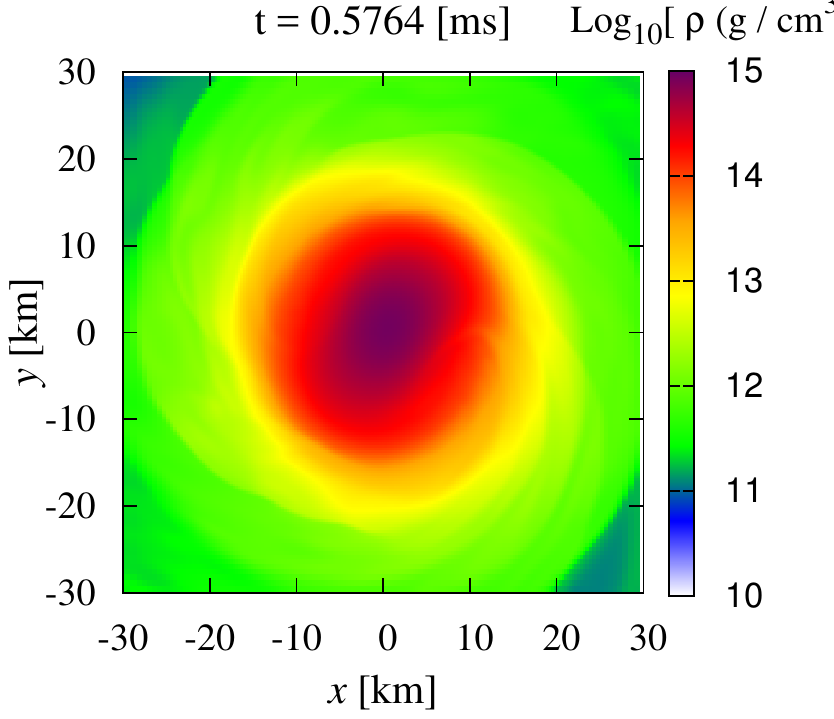} \\
\vspace{0.2cm}
\includegraphics[width=56mm]{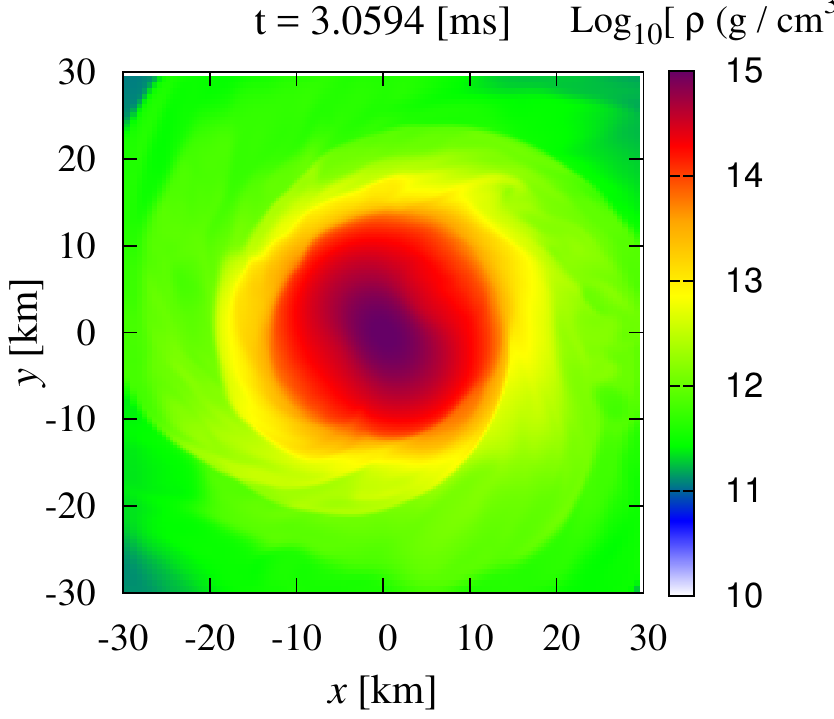}
\includegraphics[width=56mm]{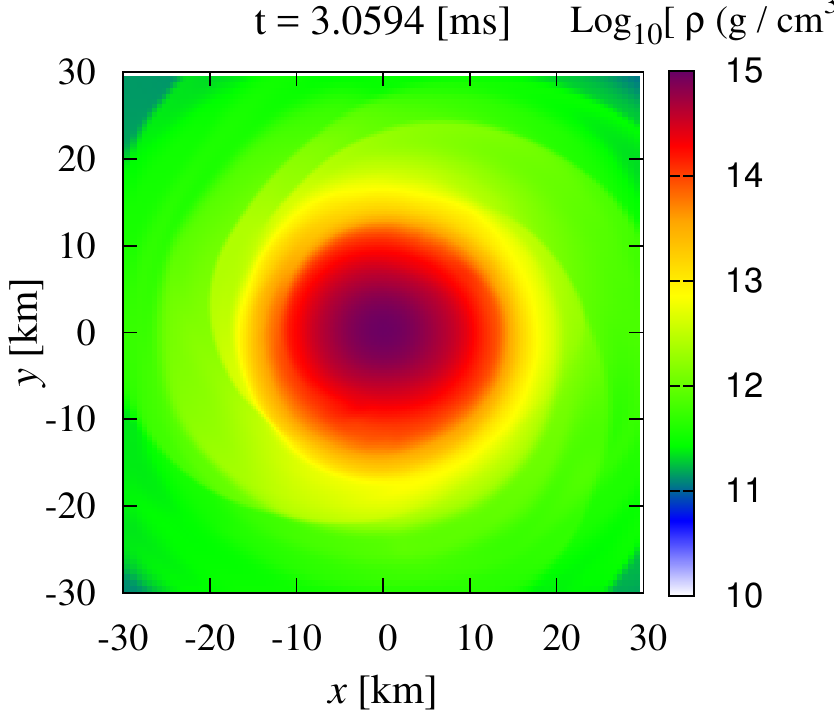}
\includegraphics[width=56mm]{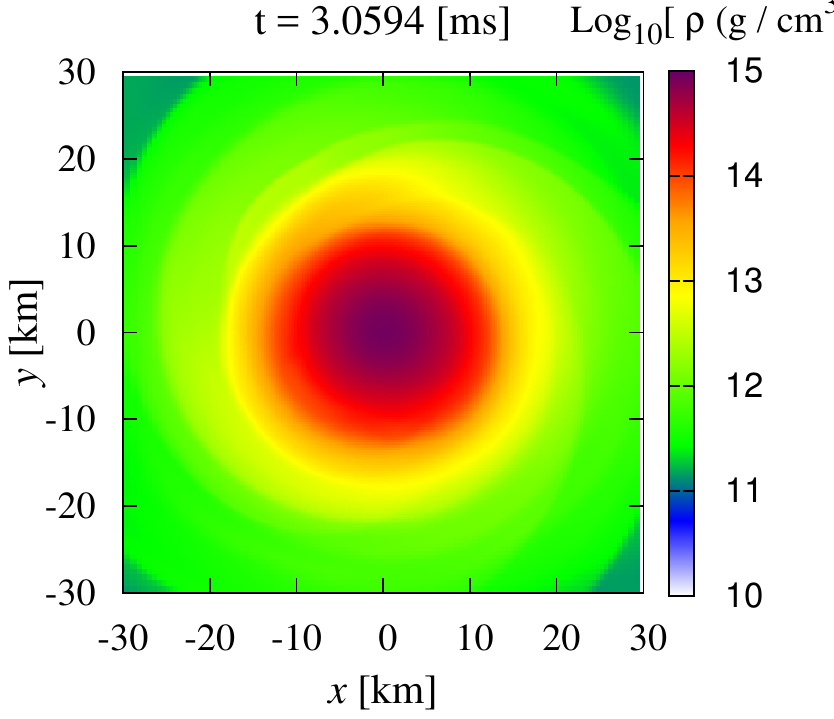} \\
\vspace{0.2cm}
\includegraphics[width=56mm]{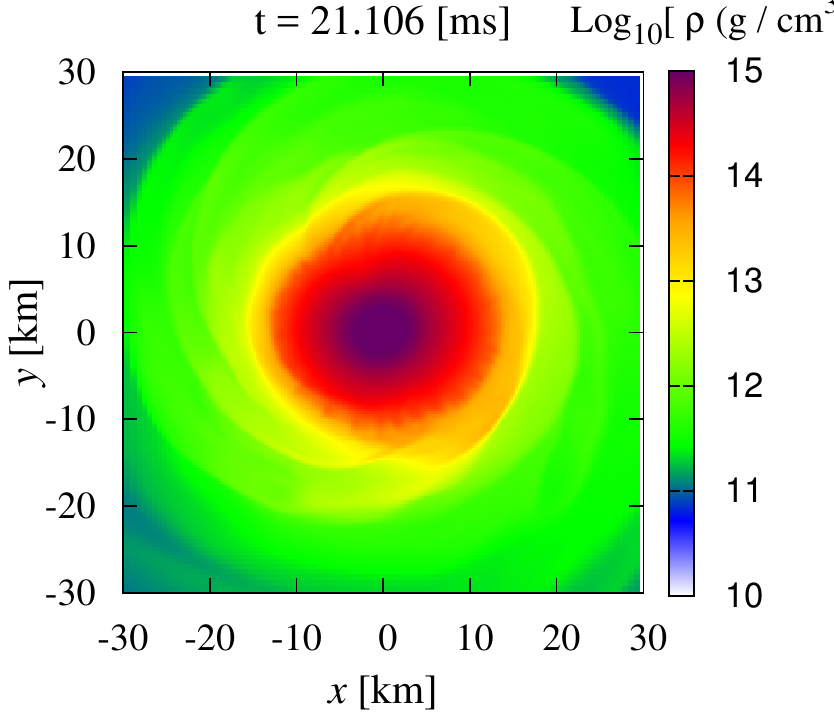}
\includegraphics[width=56mm]{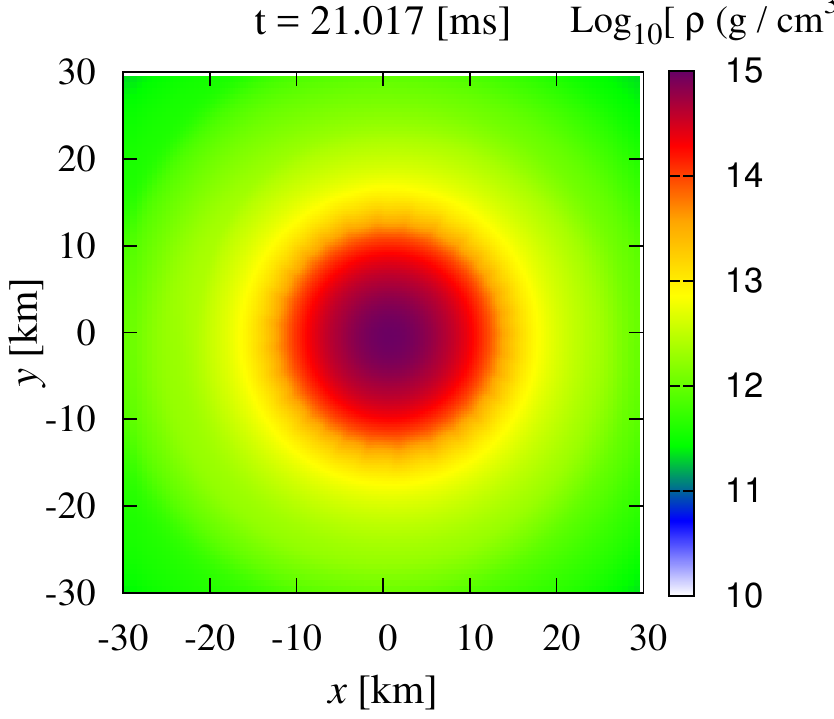}
\includegraphics[width=56mm]{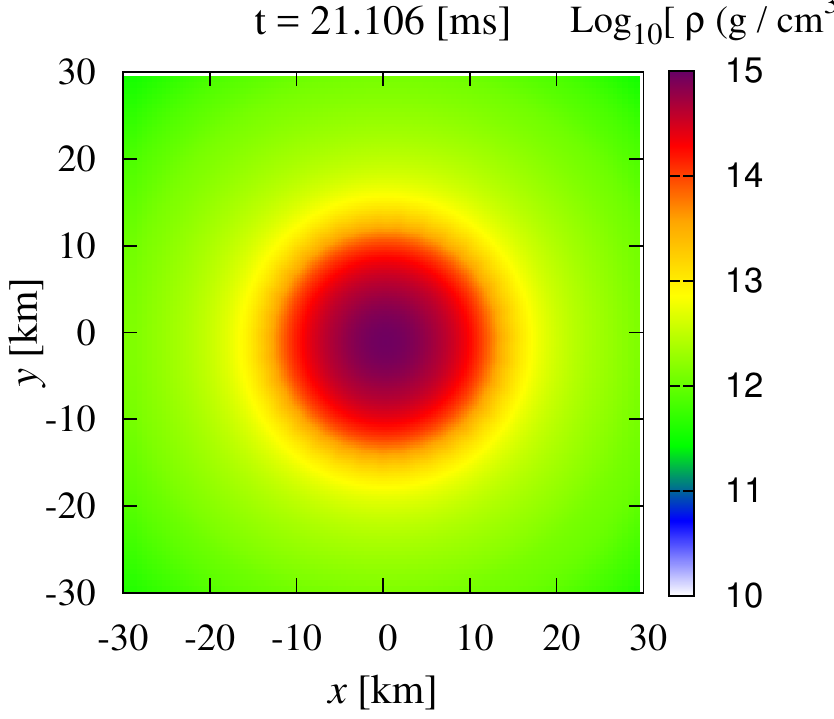} 
\caption{Evolution of rest-mass density profiles on the equatorial plane with
  $\alpha_v=0$ (left column), 0.01 (middle column), and 0.02 (right
  column).  Each column shows three snapshots of different time: The
  top, middle, and bottom rows show the results for $t \approx 0$, 3,
  and 21\,ms, respectively.
  \label{fig1}
}
\end{center}
\end{figure*}

\begin{figure*}[t]
\begin{center}
\includegraphics[width=56mm]{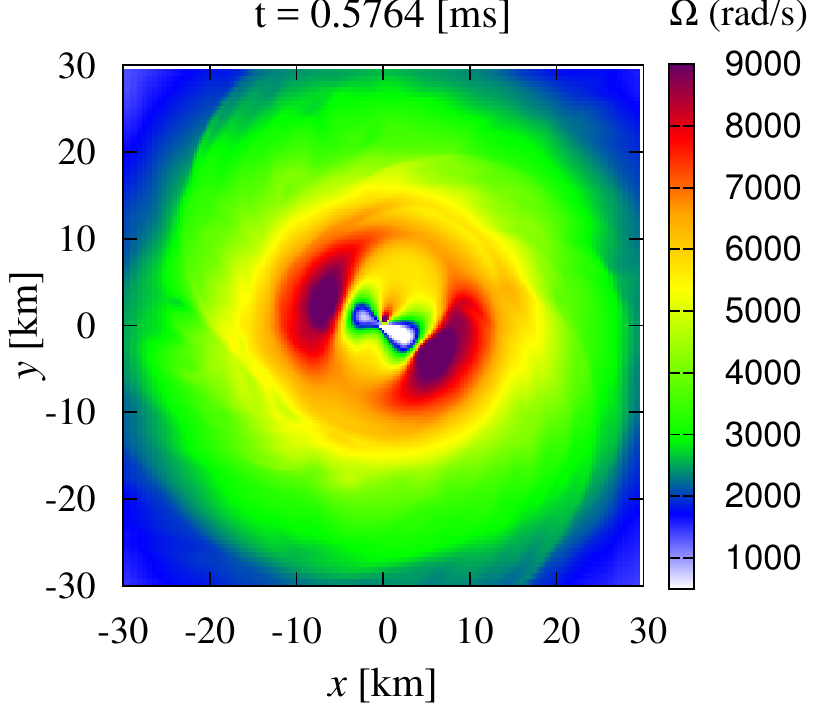}
\includegraphics[width=56mm]{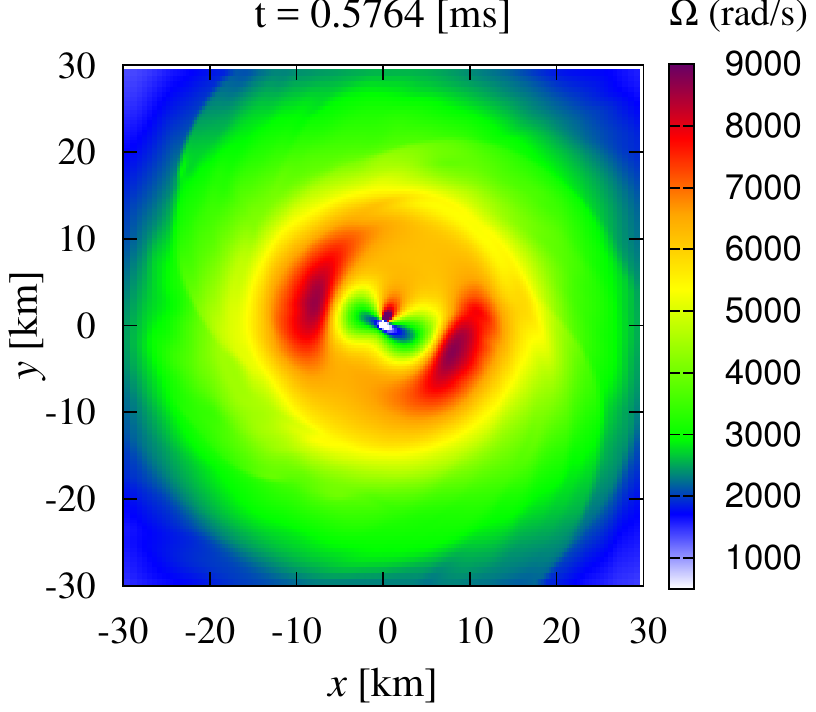}
\includegraphics[width=56mm]{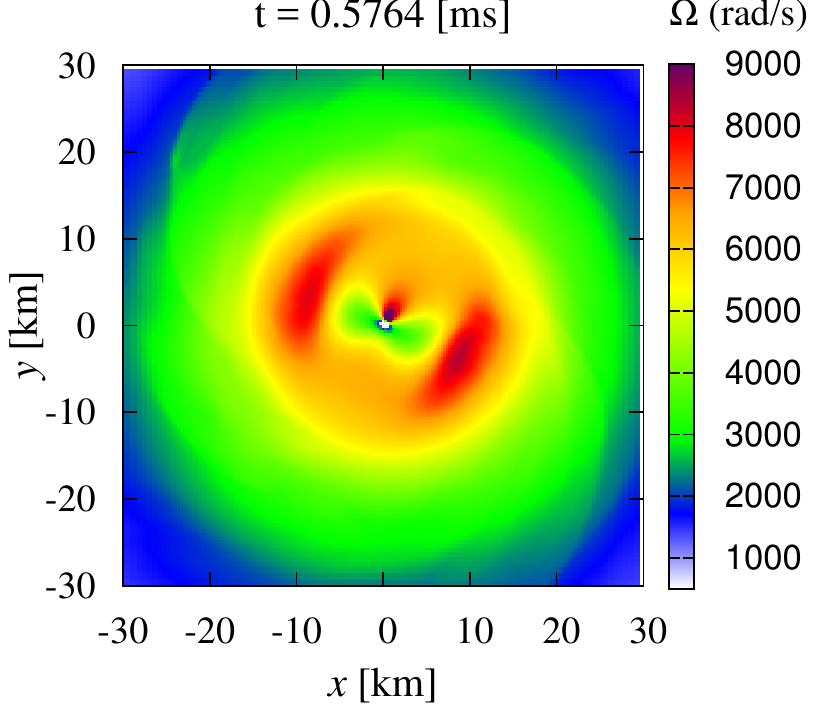} \\
\vspace{0.2cm}
\includegraphics[width=56mm]{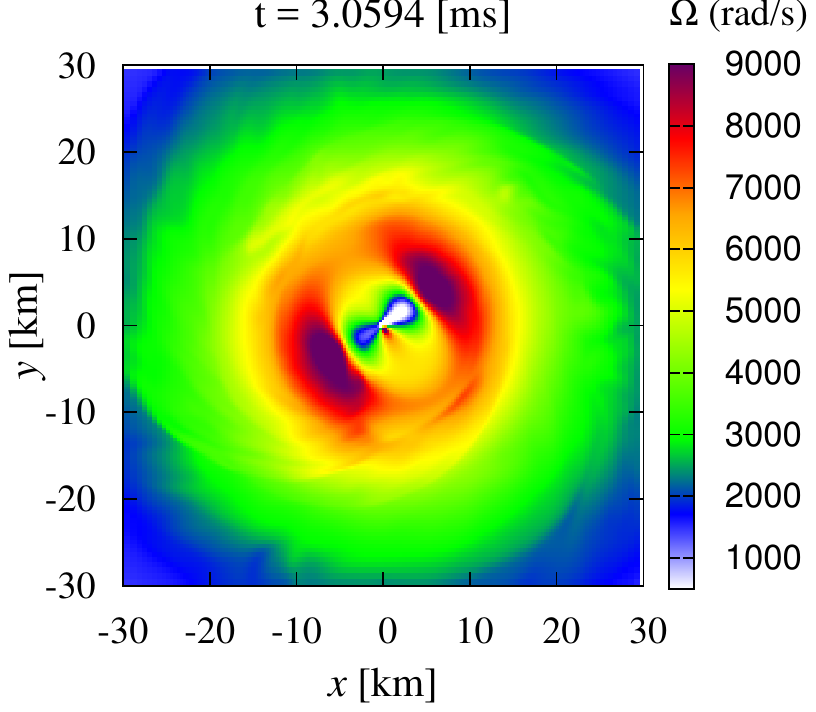}
\includegraphics[width=56mm]{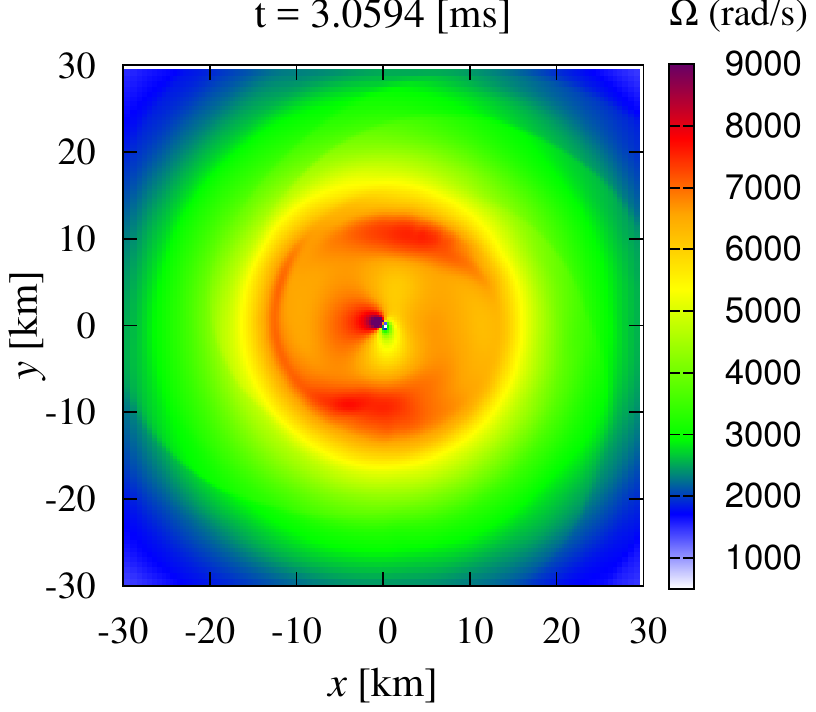}
\includegraphics[width=56mm]{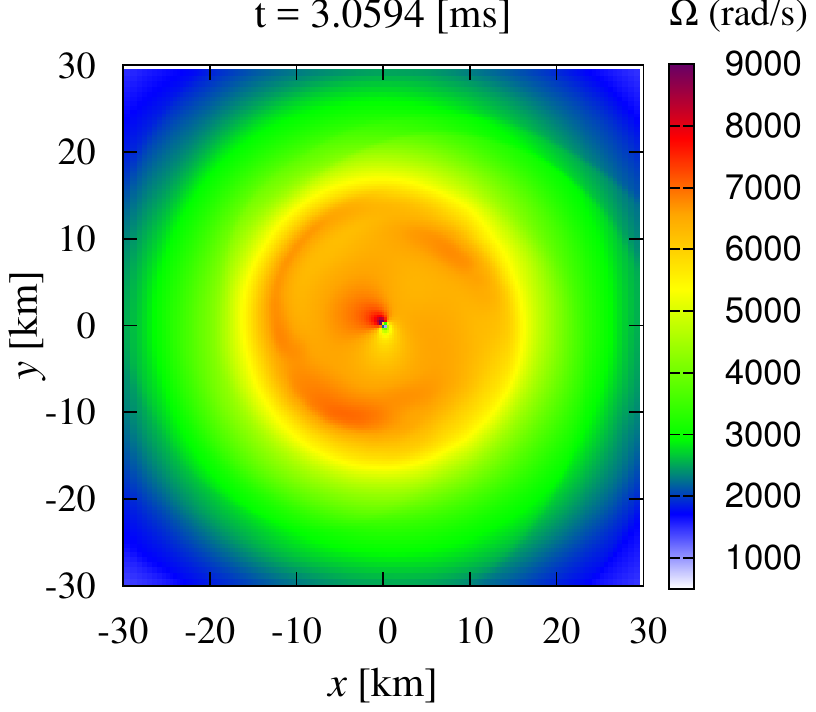} \\
\vspace{0.2cm}
\includegraphics[width=56mm]{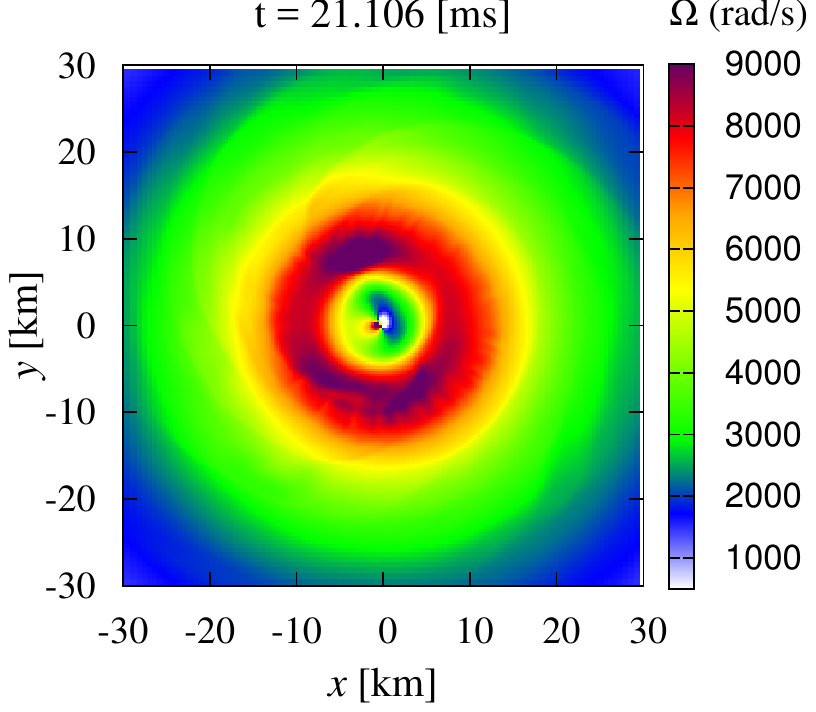}
\includegraphics[width=56mm]{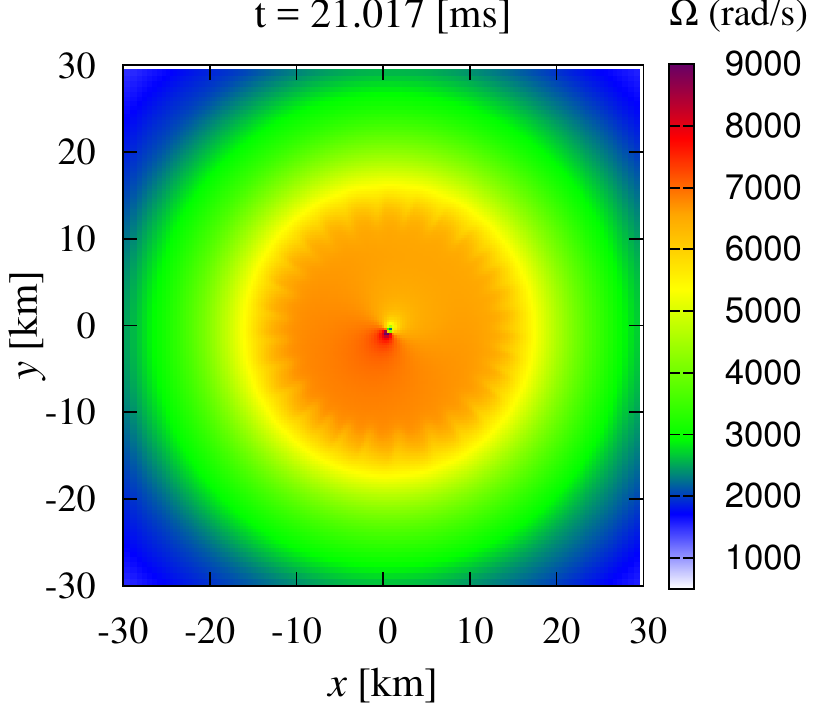}
\includegraphics[width=56mm]{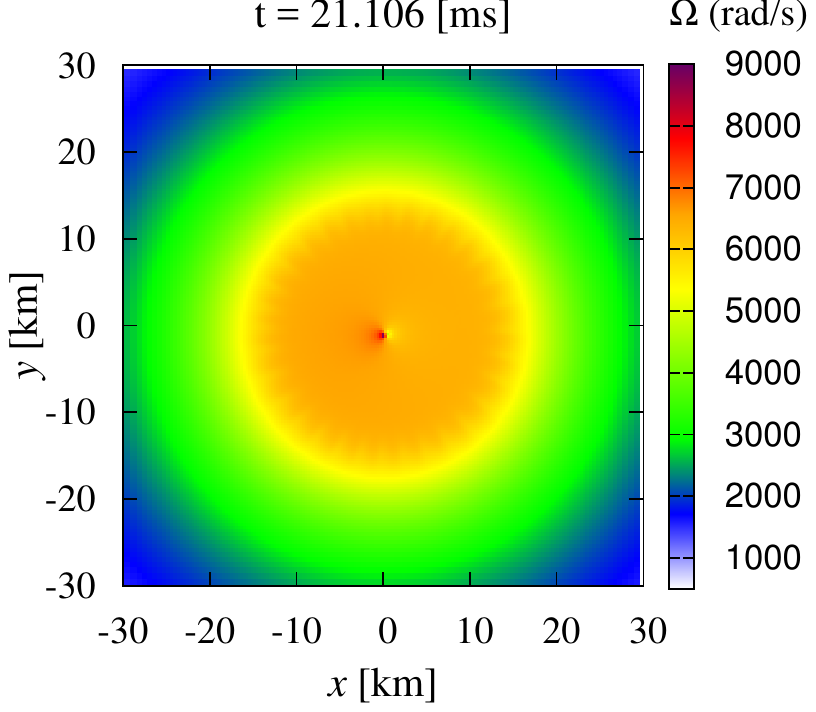} 
\caption{The same as Fig.~\ref{fig1} but for the angular velocity.
  The angular velocity is determined by
  $[-(y-y_0)v^x+(x-x_0)v^y]/[(x-x_0)^2+(y-y_0)^2]$ where $(x_0,
  y_0)$ is a mass center appropriately determined: $x_0$ and $y_0$ are
  slightly different from zero. Note that for the white region (near
  $(x_0, y_0)$), we cannot accurately determine the angular velocity.
  \label{fig2}
}
\end{center}
\end{figure*}

Figures~\ref{fig1} and \ref{fig2} display the evolution of the
profiles for the rest-mass density and angular velocity on the
equatorial plane, respectively.  For calculating the angular velocity,
we first determine the center of mass $(x_0, y_0)$ of the massive
neutron star on the equatorial plane. Here, $x_0$ and $y_0$ are in
general slightly different from zero.  Then, the angular velocity is
defined by $\Omega=[-(y-y_0)v^x+(x-x_0)v^y]/R^2$ where
$R=\sqrt{(x-x_0)^2+(y-y_0)^2}$.  For these figures, the left, middle,
and right columns show the results for $\alpha_v=0$, 0.01, and 0.02,
respectively.  The top, middle, and bottom rows show the results for
$t \approx 0$, 3, and 21\,ms, respectively.

As we have always observed in the merger
simulations~\cite{Hotokezaka2011,bauswein,Hotokezaka2013,takami,tim,ms16,luca},
the remnants of binary neutron star mergers have a non-axisymmetric
structure.  In the absence of viscous effects, this non-axisymmetric
structure is preserved for the timescale of $> 10$\,ms (see the left
column of Fig.~\ref{fig1}) because of the absence of efficient
processes of angular momentum transport, and thus, the differential
rotation is preserved (see the left column of Fig.~\ref{fig2}).  Note
that the torque exerted by the massive neutron star of
non-axisymmetric structure to surrounding material transports angular
momentum outwards, but its timescale is not as short as 10\,ms.

By contrast, in the presence of viscosity with $\alpha_v=O(10^{-2})$,
the angular momentum transport process works efficiently.  The middle
and right columns of Fig.~\ref{fig2} illustrate that the angular
velocity profile is modified approximately to a uniform profile for $R
< 15$\,km in the timescale of less than 10\,ms. As a result, the
rest-mass density profile is also modified quickly.  The middle and
right columns of Fig.~\ref{fig1} illustrate that in a timescale of
$\ll 20$\,ms, the non-axisymmetric structure disappears and the
remnant becomes approximately axisymmetric.  The disappearance of the
non-axisymmetric structure is reflected in the fact that spiral
arms in the envelope disappear for $\alpha_v=0.01$ and 0.02 (by
contrast, for $\alpha_v=0$, the spiral arms are continuously
observed). This results from the viscous effect by which torque
exerted by the remnant massive neutron star is significantly weaken
due to the reduced degree of the non-axisymmetry.  It is also found
from Fig.~\ref{fig1} that in the presence of nonzero viscosity, the
matter expands outwards: For $\alpha_v=0$, the region with $\rho \geq
10^{12}\,{\rm g/cm^3}$ is extended only to $R \sim 20$\,km but for
$\alpha_v=0.01$ and 0.02, it is extended to $R \sim 30$\,km.

\begin{figure*}[t]
\begin{center}
\includegraphics[width=56mm]{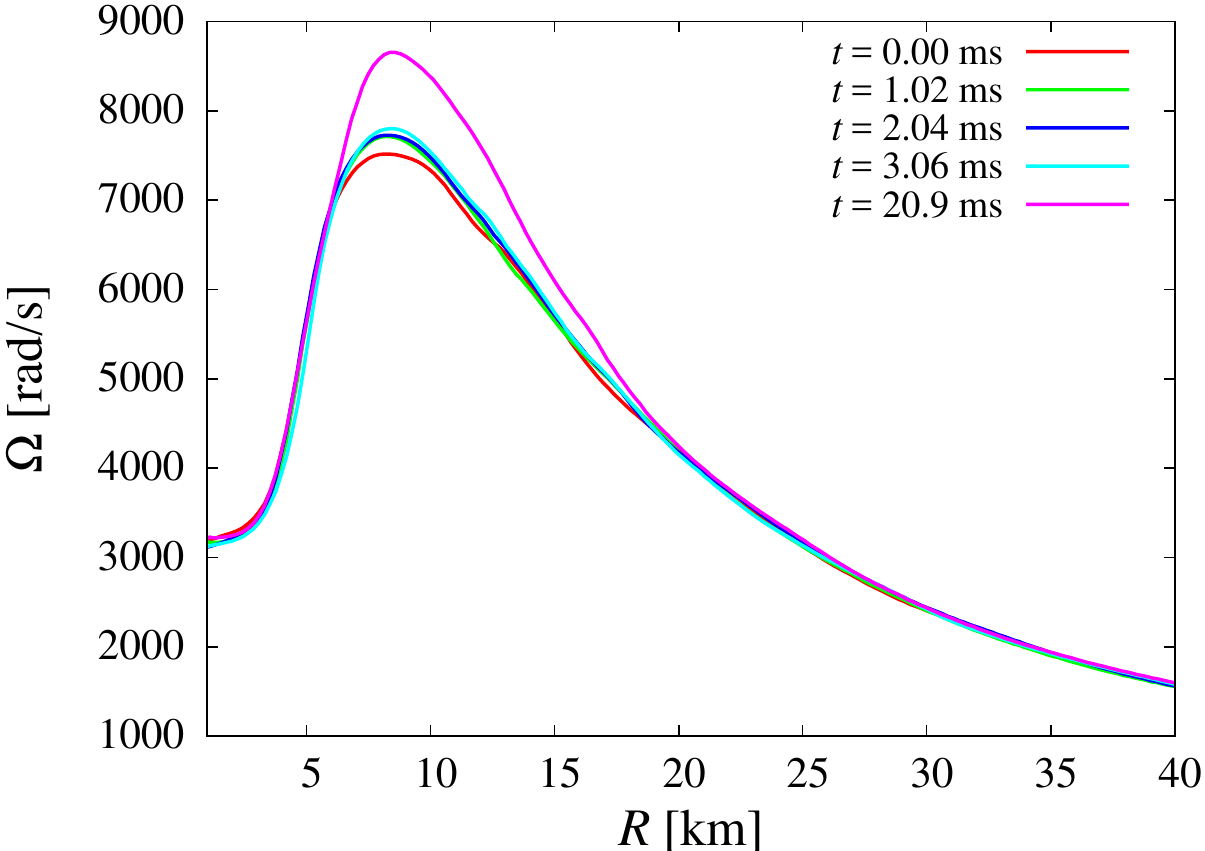}~
\includegraphics[width=56mm]{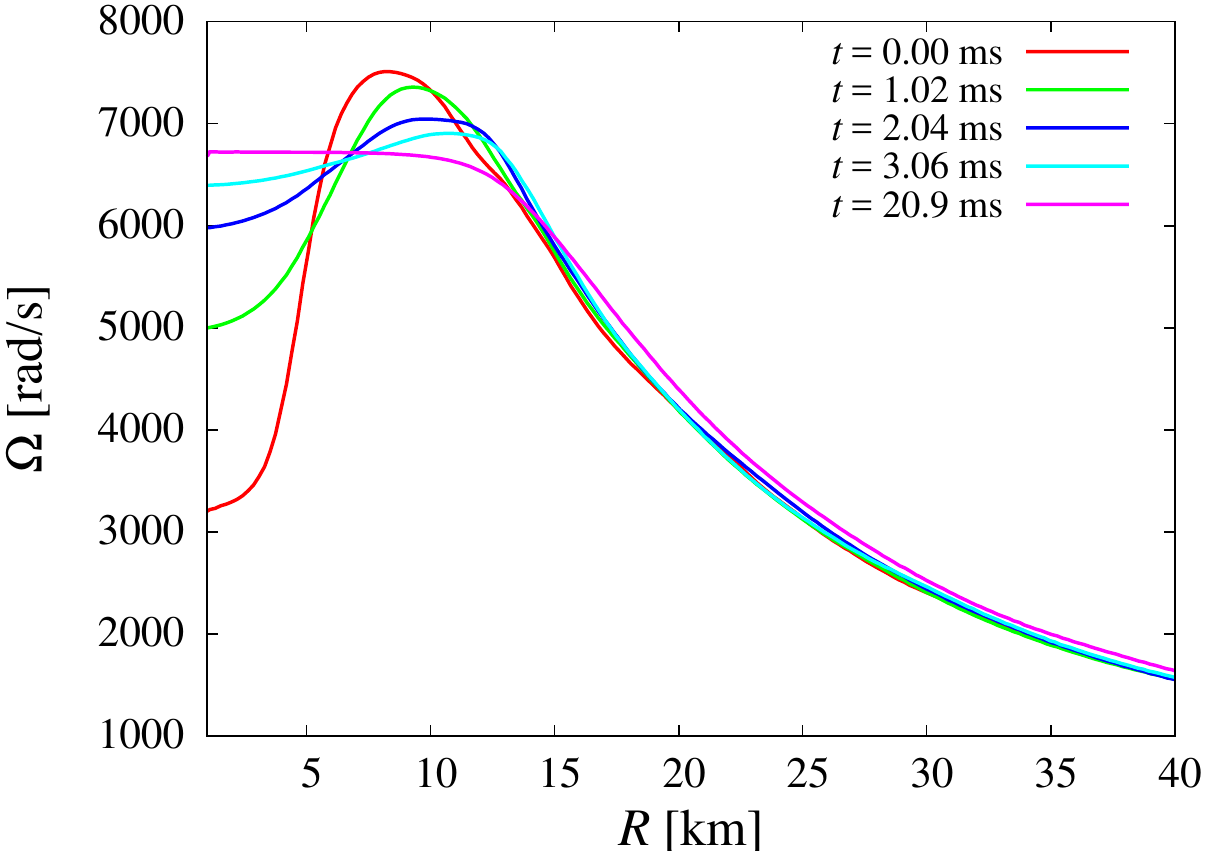}~
\includegraphics[width=56mm]{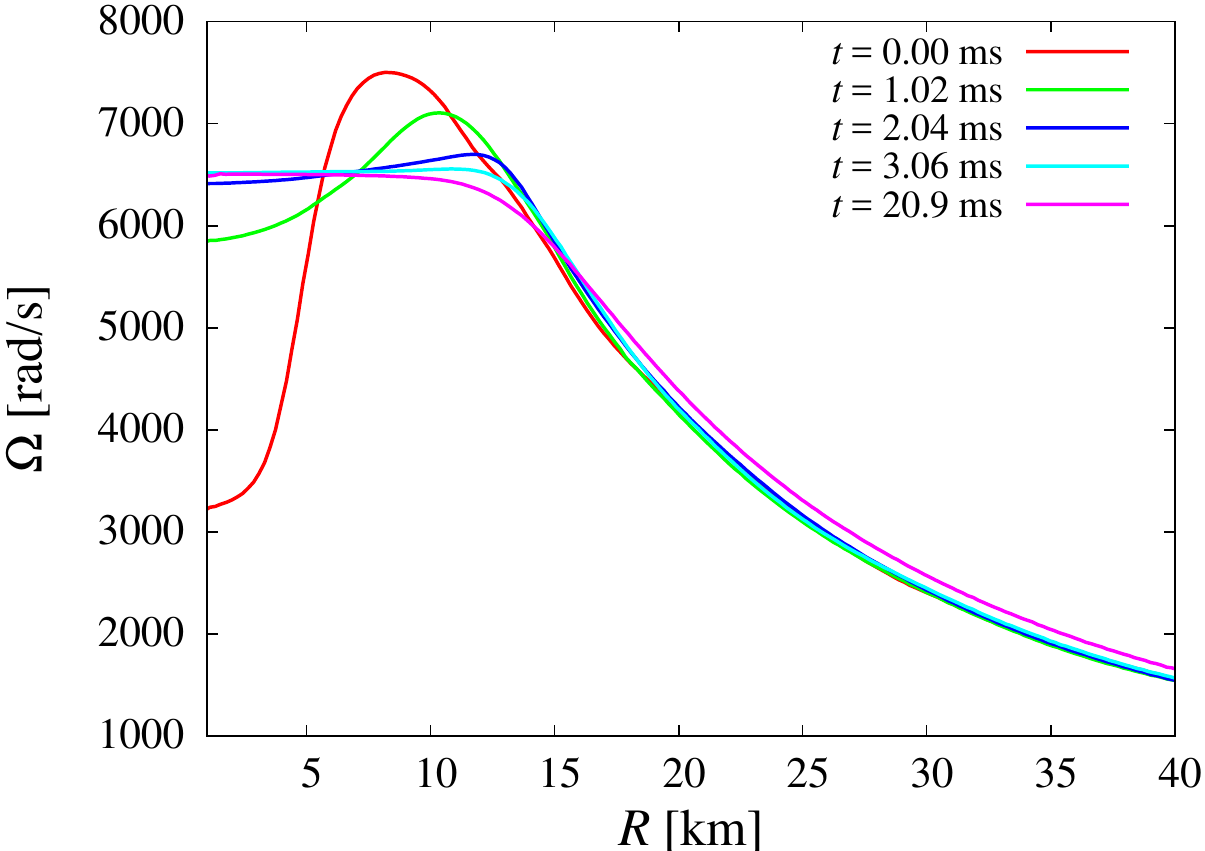}
\caption{Averaged angular-velocity profiles along the cylindrical
  radius at selected time slices for the models with $\alpha_v=0$ (left
  column), 0.01 (middle column), and 0.02 (right column).
  \label{fig3}
}
\end{center}
\end{figure*}

% Fig4

\begin{figure*}[t]
\begin{center}
\includegraphics[width=84mm]{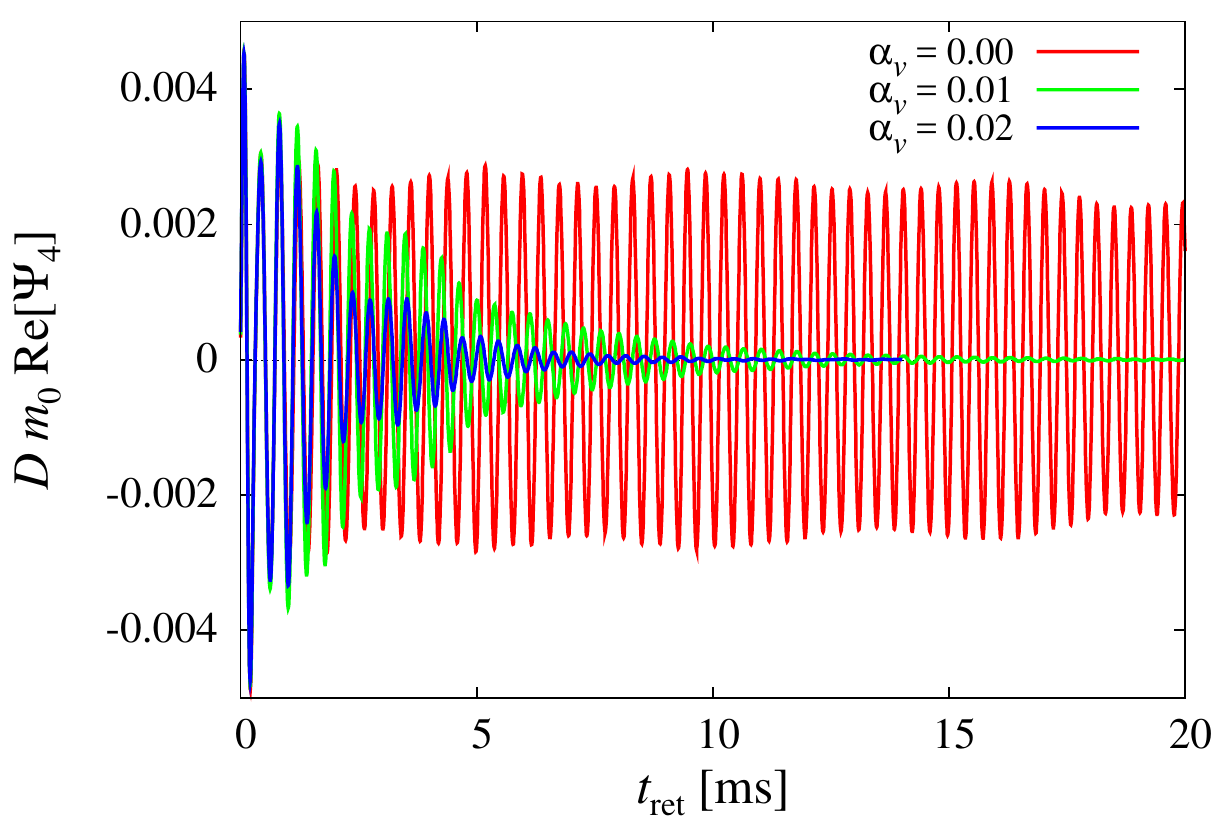}~~~~
\includegraphics[width=84mm]{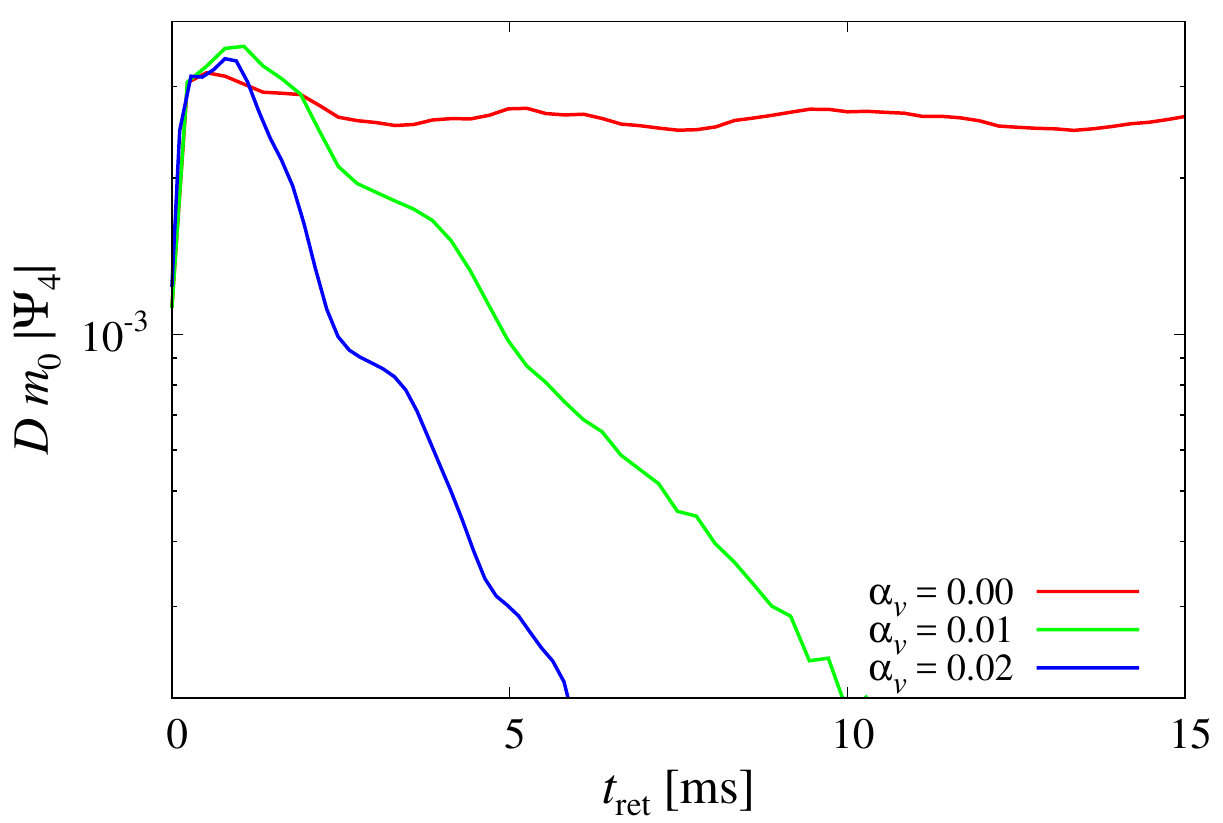}
\caption{Left: Gravitational waveforms for $\alpha_v=0$, 0.01, and
  0.02. Here we plot the real part of $\Psi_4$ for the $l=m=2$
  spin-weighted spherical-harmonics component.  $D$ and
  $m_0(=2.7M_\odot)$ are the distance from the source to the observer
  and initial gravitational mass of the binary system. We assume that
  the observer of gravitational waves is located along the rotation
  axis of the massive neutron star.  Right: Evolution of the absolute
  value of $\Psi_4$. To clarify the exponential damping of the curves,
  for plotting the right panel, we apply a filter to $|\Psi_4|$ in
  order to erase high-frequency noises.
\label{fig4}
}
\end{center}
\end{figure*}

Figure~\ref{fig3} shows the evolution of the averaged angular velocity
along the cylindrical radius $R$. The averaged angular velocity is
obtained by averaging $\Omega$ along the azimuthal angle
$\varphi=\tan^{-1}[(y-y_0)/(x-x_0)]$ for fixed values of $R$ and
$z$ as
\beq
\bar\Omega(R,z) = {1 \over 2\pi} \int_0^{2\pi} \Omega(R,z,\varphi)\,d\varphi .
\eeq

As often observed in numerical relativity simulations~(e.g.,
Ref.~\cite{STU2005}), the merger remnants are initially differentially
rotating, and the maximum of the angular velocity, $\Omega$, is
located in the vicinity of the stellar surface of the remnant massive
neutron stars. Such profiles are formed during the merger process,
because at the onset of the merger, two neutron stars with opposite
velocity vectors collide with each other and kinetic energy is
dissipated at their contact surfaces. In the absence of the viscous
effects, this profile is preserved for a timescale longer than 20\,ms.
By contrast, in the presence of the viscous effects, this differential
rotation disappears in the viscous timescale, $t_{\rm vis}$, because
of the efficient viscous angular momentum transport. Because the peak
of $\Omega$ is initially located near the stellar surface, the angular
velocity near the rotation axis is increased. That is, the angular
momentum is transported inward inside the massive neutron star.  In
this example, the rotation period of the resulting massive neutron
star, $2\pi/\Omega$, is relaxed uniformly to be $\approx 1$\,ms.

%% ADD

Associated with the viscous angular momentum transport, a massive
torus surrounding the central massive neutron star develops in the
presence of viscosity. The rest mass of the torus, measured for
the matter with $R \geq 10$ and 15\,km at $t=20$\,ms, is $\approx
0.41$ and $0.19M_\odot$ for $\alpha_v=0.02$ and $\approx 0.38$ and
$0.17M_\odot$ for $\alpha_v=0.01$ , respectively. The torus mass for
$\alpha_v=0$ at $t=20$\,ms is $\approx 0.26$ and $0.15M_\odot$ for $R
\geq 10$ and 15\,km, and hence, the viscous angular momentum transport
enhances the massive torus formation.

As shown in Fig.~\ref{fig3}, the torus is preserved to be
differentially rotating around the central massive neutron star, and
hence, it is still subject to viscous heating and viscous angular
momentum transport.  As indicated in our previous study~\cite{SKS16},
the inner part of the torus will be heated up significantly due to the
long-term viscous heating, and eventually, an outflow could be driven
with a substantial amount of mass ejection. Exploring the generation
processes of the outflow and mass ejection by a long-term
numerical-relativity simulation is one of our future issues.

%%%%%%%%%%%%%%%%

These viscous effects are reflected intensely in gravitational waves
emitted by the remnant massive neutron star. Figure~\ref{fig4} shows
the gravitational waveforms for $\alpha_v=0$, 0.01, and 0.02 all
together. Here, we plot the real part of the out-going components of
the complex Weyl scalar (the so-called $\Psi_4$) for the $l=m=2$
spin-weighted spherical-harmonics component. For $\alpha_v=0$,
quasi-periodic gravitational waves are emitted for a timescale much
longer than 10\,ms, reflecting the fact that the non-axisymmetric
structure of the massive neutron star is preserved (this waveform is
essentially the same as that we found in our merger
simulation~\cite{Hotokezaka2013}).  The right panel of Fig.~\ref{fig4}
clearly shows that the amplitude of gravitational waves is nearly
constant for this case.  By contrast, the gravitational-wave amplitude
decreases exponentially in time in the viscous timescale for
$\alpha_v\not=0$. This reflects the fact that the non-axisymmetric
structure of the massive neutron star is damped in the viscous
timescale.  This suggests that, in the presence of MHD turbulence
which would induce turbulence viscosity, the merger remnant of binary
neutron stars may not emit high-amplitude gravitational waves for a
timescale longer than $\sim 10$\,ms.

The right panel of Fig.~\ref{fig4} shows that the amplitude of
gravitational waves damps in an exponential manner $\propto
\exp(-t/\tau)$ where $\tau$ is the $e$-holding damping timescale. In
the present results, $\tau$ is approximately written as
\beq
\tau \approx 3.6 \,{\rm ms}\left({\alpha_v \over 0.01}\right)^{-1}.
\label{eq:tau1}
\eeq
This timescale agrees approximately with the timescale for the change of
the angular velocity found in Fig.~\ref{fig3}. 

We note that the author of Ref.~\cite{radice} recently performed a
viscous hydrodynamics simulation for the merger of binary neutron
stars. He also showed that the luminosity of gravitational waves
decreases with the increase of the viscous coefficient, although he
focused only on the case with a small viscous parameter (in the
terminology of alpha viscosity, he focuses only on the cases of
$\alpha_v = O(10^{-3})$ or less). 

\begin{figure}[t]
\begin{center}
\includegraphics[width=84mm]{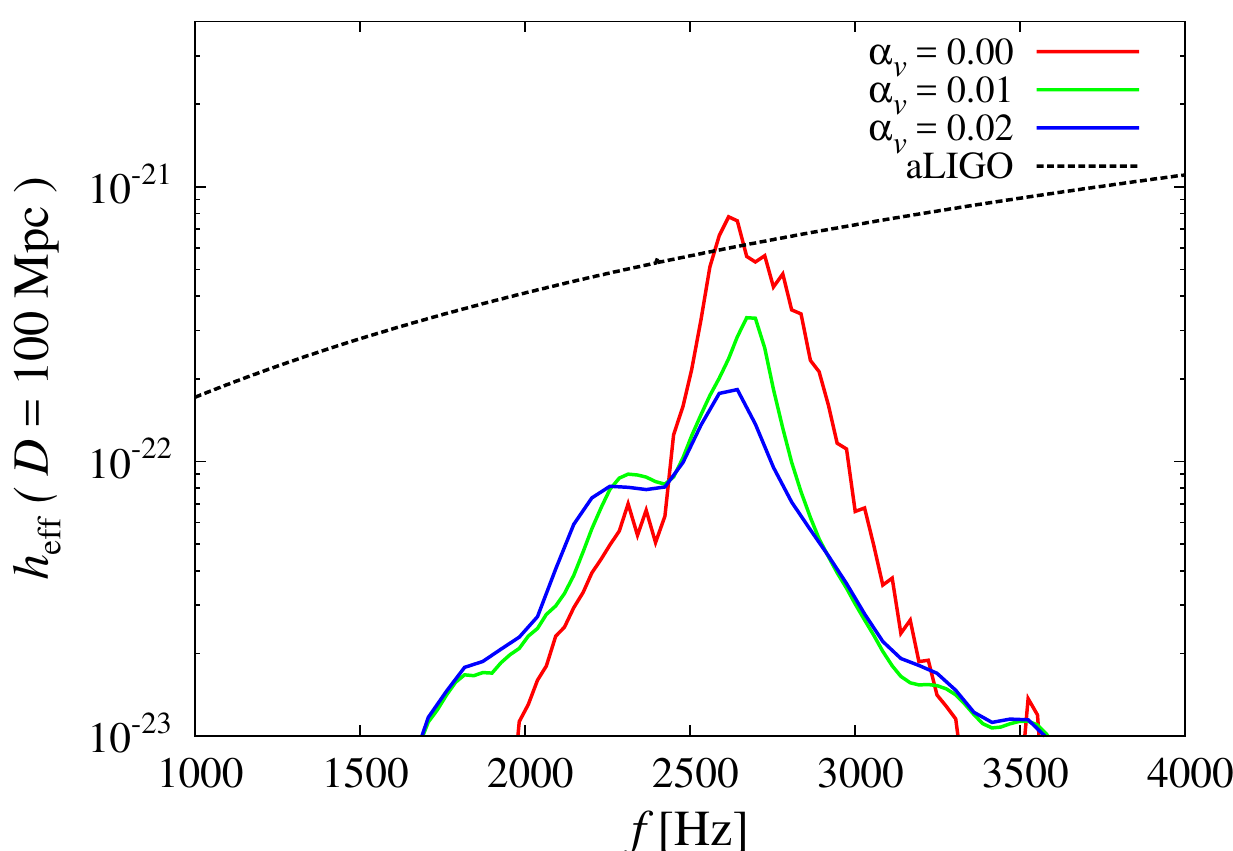}
\caption{The spectrum of gravitational waves for $\alpha_v=0$, 0.01,
  and 0.02 with $D=100$\,Mpc. We assume that the observer is located
  along the most optimistic direction (i.e., along the rotation axis).
  The spectrum, $h_{\rm eff}$, is defined by $|\tilde h(f)f|$ where
  $\tilde h(f)$ is the Fourier spectrum of gravitational waves. The
  dashed curve referred to as aLIGO denotes $[S_n(f) f]^{1/2}$ where
  $S_n$ is the one-sided noise spectrum density for the ``Zero
  Detuning High Power'' configuration of Advanced
  LIGO~\cite{ligonoise}.
  \label{fig5}
}
\end{center}
\end{figure}

Figure~\ref{fig5} plots the spectrum of gravitational waveforms
defined by $h_{\rm eff}:=|f \tilde h(f)|$ where $\tilde h(f)$ is the
Fourier transformation of gravitational waves defined by
\beqn
\tilde h(f):=\int_0^T h(t) e^{-2\pi i f t} dt. 
\eeqn
Here $T$ corresponds to the final time of the simulation and $h(t)=h_+ -
i h_\times$ with $h_+$ and $h_\times$ denoting the plus and cross
modes of gravitational waves. In numerical simulation, we obtain
$\ddot h(t)(=\Psi_4)$, and hence, we calculate $\tilde h(f)$ by
\beqn
\tilde h(f)=-\int_0^T {\Psi_4(t) \over (2\pi f)^2} e^{-2\pi i f t} dt.
\eeqn

Figure~\ref{fig5} shows that the peak amplitude decreases
monotonically with the increase of $\alpha_v$, while the peak
frequency depends only weakly on the value of $\alpha_v$.  We note
that we stopped our simulations at $t \approx 25$\,ms, and hence, for
$\alpha_v=0$ for which gravitational waves are emitted for a timescale
longer than 25\,ms, the peak amplitude of $\tilde h(f)$ would be
underestimated. 

To qualitatively understand the behavior on the decrease of the peak
amplitude for the spectrum, we consider a simple model in which the
gravitational waveforms denoted by $h:=h_+ - i h_\times$ are written
as a monochromatic form $A\exp(2\pi i f_p t-\tau_v t)$ where $A$ is
the amplitude of $h$, $f_p$ is the monochromatic frequency of
gravitational waves, and $\tau_v$ denotes the damping time of the
gravitational-wave amplitude (which should be proportional to
$\alpha_v$ and approximately equal to $\tau$ in
Eq.~(\ref{eq:tau1})). Then, at the peak of the Fourier spectrum,
$f=f_p$, we obtain
\beqn
|\tilde h(f_p)|={1 - \exp(-\tau_v T) \over \tau_v}A. 
\eeqn
Thus, for $\tau_v T \gg 1$, $|\tilde h(f_p)|=A\tau_v^{-1}$.  Since
$|\tilde h(f_p)|=AT$ for $\tau_v \rightarrow 0$, the peak amplitude is
by a factor of $(\tau_v T)^{-1}$ (i.e., by a factor of several)
smaller than that in the non-viscous case.  The numerical results
agree qualitatively with this rule.

\section{Summary}

Employing a simplified formalism for general relativistic viscous
hydrodynamics that can minimally capture the effects of the shear
viscous stress, we performed numerical-relativity simulations for the
evolution of a remnant of binary neutron star merger, paying
particular attention to gravitational waves emitted by the merger
remnant. As often found in the purely hydrodynamical
simulations~\cite{STU2005,Hotokezaka2011,bauswein,Hotokezaka2013,takami,tim,ms16,luca},
in the absence of viscous effects, the merger remnant emits
quasi-periodic gravitational waves for a timescale longer than 10\,ms
keeping their amplitude high and their frequency approximately
constant. However, in the presence of viscous effects, the amplitude
of gravitational waves damps in an exponential manner in time. The
timescale of the exponential damping agrees approximately with the
viscous timescale, which is $\alt 5$\,ms for a plausible viscous
coefficient with $\alpha_v \agt 0.01$.  This suggests that the merger
remnant of binary neutron stars may not be a strong emitter of
gravitational waves.

In reality, the viscous effects would be induced by MHD turbulence.
Since we do not know whether the MHD turbulence is equivalent to the
shear viscous-hydrodynamical turbulence, our present result is still
speculative. To obtain the true answer in this problem, we have to
perform an extremely-high-resolution MHD simulation in future.
However, it will not be an easy task to perform such simulations in
the near future because of the limitation of computational resources.
Thus, the best attitude we can currently take is to keep in mind that
there would be a variety of the possibilities for the evolution of the
merger remnants of binary neutron stars because we have not yet fully
understood the physics for the merger remnant.  Thus, even if
gravitational waves of a significant amplitude are not observed for
the post merger phase of binary neutron star mergers in the
near-future gravitational-wave observation, we should not consider
that such observation implies that a massive neutron star is not
formed after the merger. On the other hand, if gravitational waves of
a significant amplitude are observed for the post merger phase of
binary neutron star merger, we will be able to conclude that a massive
neutron star is formed as a remnant and viscous effects do not play an
important role for the merger remnant. 

%Before closing this paper, we note that during our final preparation
%phase of this paper, David Radice~\cite{radice} submitted his paper to
%arXiv. He describes another viscous hydrodynamics formalism that works
%well. Although he focuses only on the case with a small viscous
%parameter (in the terminology of alpha viscosity, he focuses only on
%the cases of $\alpha_v = O(10^{-3})$ or less), we find that his
%results agree qualitatively with our findings.

\begin{figure*}[t]
\begin{center}
\includegraphics[width=84mm]{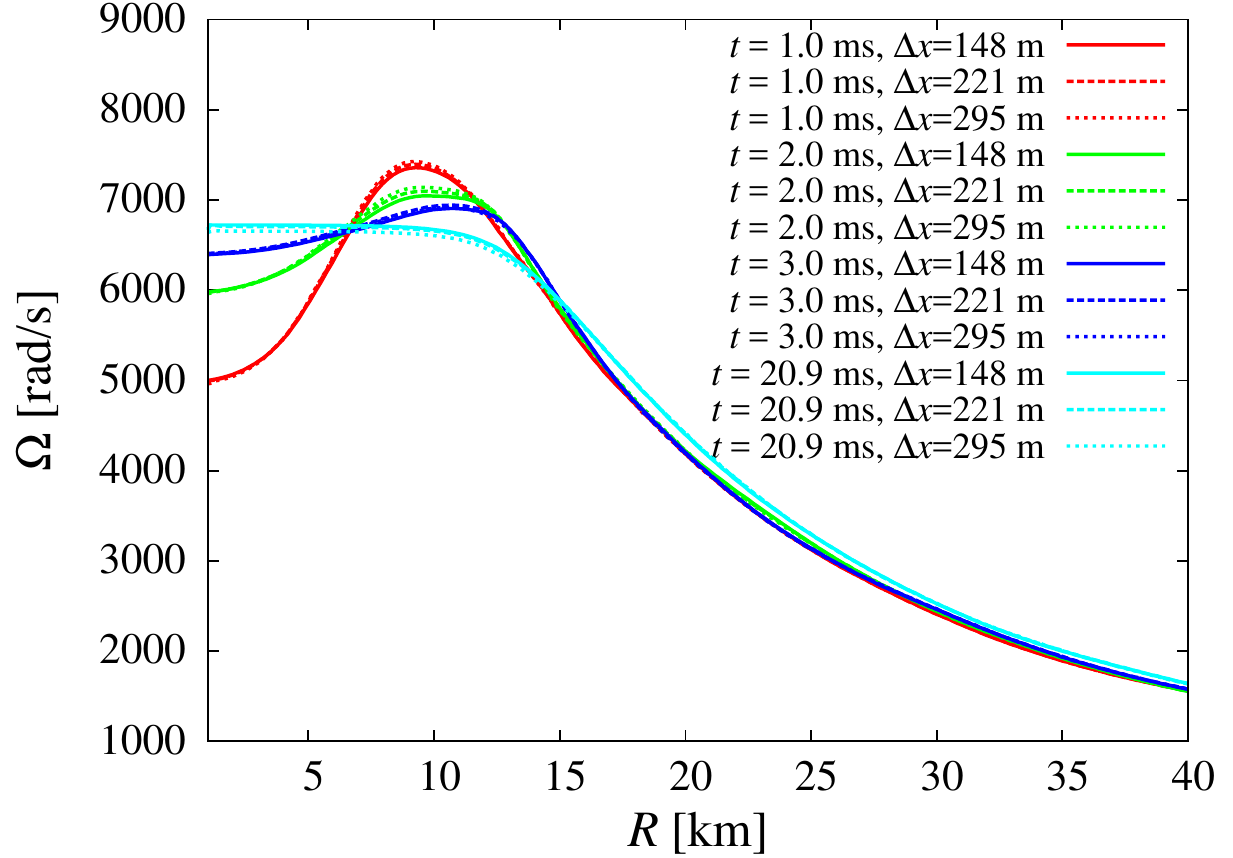}~~~
\includegraphics[width=84mm]{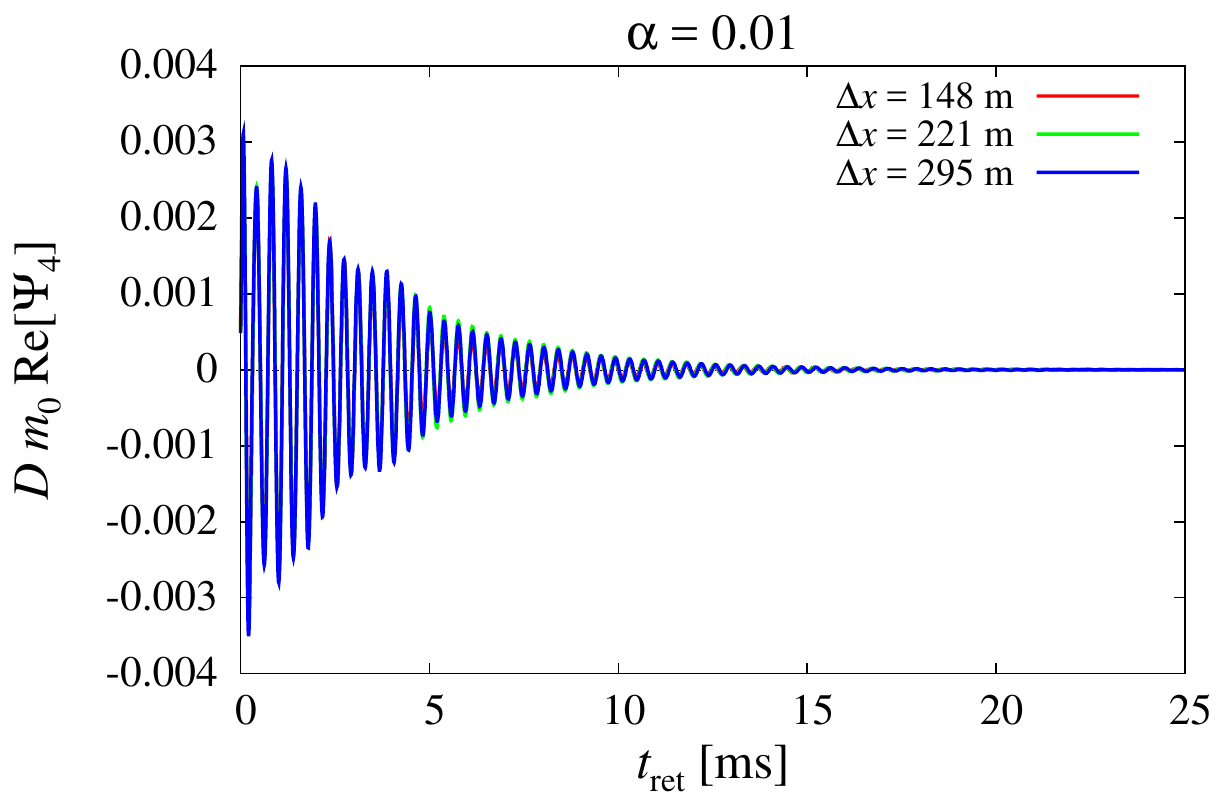}
\caption{Left: The averaged angular velocity along the cylindrical
  radius at selected time slices for three different grid
  resolutions. Right: The gravitational waveforms for $\alpha_v=0.01$
  with three different grid resolutions.
\label{fig6}
}
\end{center}
\end{figure*}

\begin{figure}[t]
\begin{center}
\includegraphics[width=84mm]{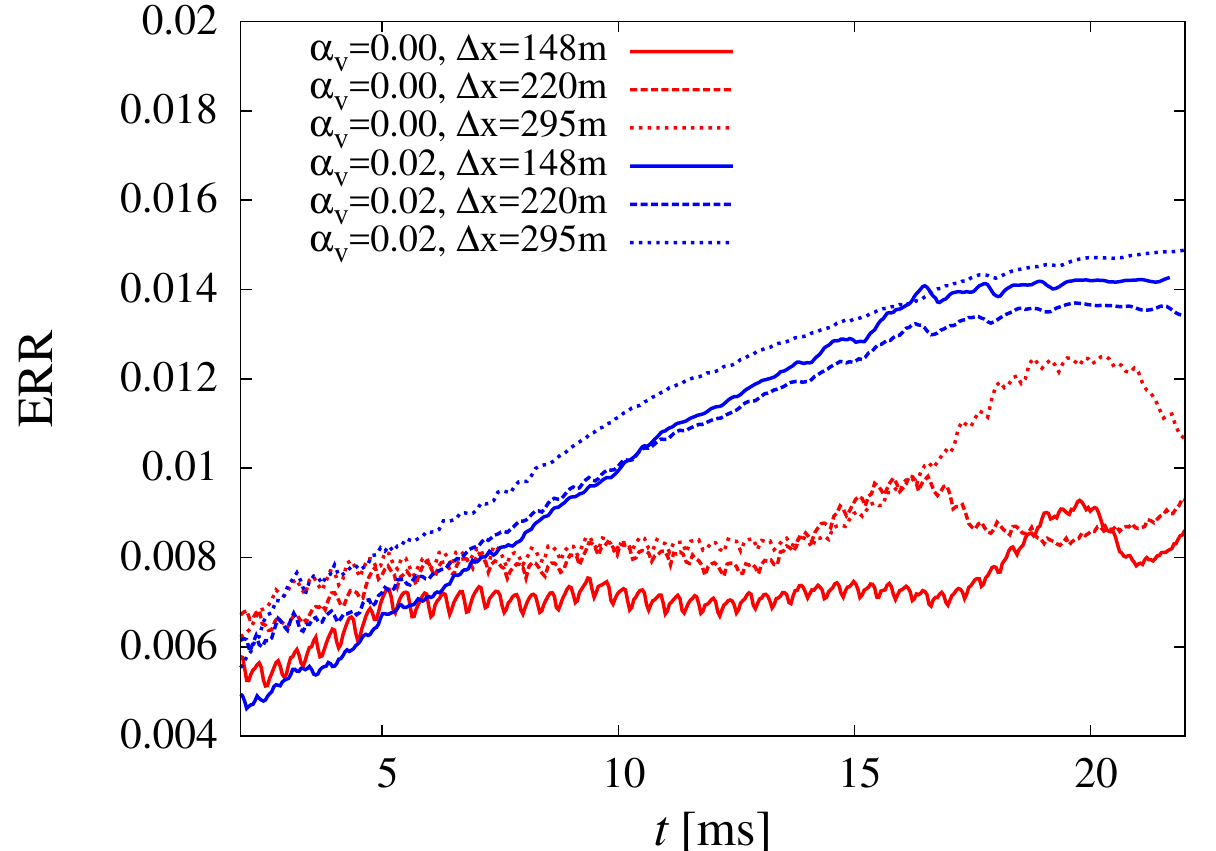}
\caption{Evolution for the violation of the Hamiltonian constraint
  (ERR defined in Eq.~(\ref{eqERR})). 
\label{fig7}
}
\end{center}
\end{figure}

\begin{acknowledgments}

This work was supported by Grant-in-Aid for Scientific Research (Grant
Nos. 24244028, 15H00783, 15H00836, 15K05077, 16H02183) of Japanese
JSPS and by a post-K computer project (Project No.~9) of Japanese
MEXT. Numerical computation was performed on Cray XC30 at cfca of
National Astronomical Observatory of Japan and on Cray XC40 at
Yukawa Institute for Theoretical Physics, Kyoto University. 

\end{acknowledgments}

\appendix

\section{Convergence and accuracy}

The results shown in this paper depend only weakly on the grid
resolution. To demonstrate this fact, we plot the averaged angular
velocity profile and gravitational waveforms in Fig.~\ref{fig6}. This
shows that the numerical results indeed depend very weakly on the grid
resolutions.  The plots of Fig.~\ref{fig6} show that our numerical
results are reliable. This result illustrates a merit of employing
viscous hydrodynamics that we do not have to perform simulations with
extremely high resolution in this framework, by contrast to the cases
of MHD simulations.

To check the accuracy of our simulations, we also monitored the
violation of the Hamiltonian constraint, denoted by $H=0$. Following
our previous paper~\cite{SKS16}, we focus on the following
rest-mass-averaged quantity,
\beq
{\rm ERR}={1 \over M_*} \int \rho_* {|H| \over \sum_k |H_k|} d^3x,
\label{eqERR}
\eeq
where $H=\sum_k H_k$ and $H_k$ denotes individual components in $H$
composed of the energy density, the square terms of the extrinsic
curvature, and three-dimensional Ricci scalar. $M_*$ denotes the total
rest mass of the system.  ERR shows an averaged degree for the
violation of the Hamiltonian constraint (for ERR=0, the constraint is
satisfied, while for ERR=1, the Hamiltonian constraint is by 100\%
violated).  Figure~\ref{fig7} plots the evolution of ERR for
$\alpha_v=0$ and 0.02 and for the three grid resolutions. As we showed
in Ref.~\cite{SKS16}, the convergence with respect to the grid
resolution is lost because the spatial derivative of the velocity,
which presents in the equations of $\tau_{ij}$, introduces a
non-convergent error.  However, the degree of the violation is kept to
be reasonably small, i.e., ERR $\alt 0.015$, in our simulation time,
irrespective of the values of $\alpha_v$.  This approximately
indicates that the Hamiltonian constraint is satisfied within 1.5\%
error. Therefore, we conclude that the results obtained in this paper
are reliable at least in our present simulation time.

%%%%%%%%%%%%%%%%%%%%%%%%%%%

%%%%%%%%%%%%%%%%%%%%%

\end{document}